\documentclass[prd, twocolumn, superscriptaddress,floatfix, nofootinbib]{revtex4-2}
\usepackage[margin=1.in]{geometry}
\usepackage[utf8]{inputenc}
\usepackage{amssymb}
\usepackage{amsmath}
\usepackage{graphicx}
\usepackage{url}
\usepackage[capitalize]{cleveref}

\begin{document}

\title{A Living Review of Machine Learning for Particle Physics}

\author{Matthew Feickert}
\email{matthew.feickert@cern.ch}
\affiliation{Department of Physics, University of Illinois at Urbana-Champaign}

\date{\today}

\author{Benjamin Nachman}
\email{bpnachman@lbl.gov}
\affiliation{Physics Division, Lawrence Berkeley National Laboratory}
\affiliation{Berkeley Institute for Data Science, University of California}

\begin{abstract}
Modern machine learning techniques, including deep learning, are rapidly being applied, adapted, and developed for high energy physics.  Given the fast pace of this research, we have created a living review with the goal of providing a nearly comprehensive list of citations for those developing and applying these approaches to experimental, phenomenological, or theoretical analyses. As a living document, it will be updated as often as possible to incorporate the latest developments. A list of proper (unchanging) reviews can be found within. Papers are grouped into a small set of topics to be as useful as possible. Suggestions and contributions are most welcome, and we provide instructions for participating.
\end{abstract}

\maketitle

\section{Introduction}

Machine learning (ML) is a generic term used to describe any automated inference procedure, broadly defined.  As such, machine learning plays a key role in nearly all areas of high energy physics (HEP).  Traditionally, machine learning has been synonymous with ``multivariate techniques'', with Boosted Decision Trees as the community favorite method and TMVA~\cite{hoecker2007tmva} as the community favorite tool.  The set of methods and tools commonly used in HEP has grown significantly in recent years as a result of the deep learning revolution.

With the rapid development of research at the intersection of machine learning and HEP, it is difficult to follow the latest developments.  This is a challenge for new researchers to integrate into the field and also for seasoned practitioners to put their work into the context of the existing literature.  To help solve this challenge, we have created a review of ML for HEP with the goal of providing a nearly comprehensive list of citations for papers that develop and apply ML to experimental, phenomenological, or theoretical analyses.  In order to be comprehensive and remain useful, this review is \textit{living} in the sense that it is continuously updated and is open for community contributions.

\begin{figure}[h!]
    \centering
    \fbox{\includegraphics[width=\linewidth]{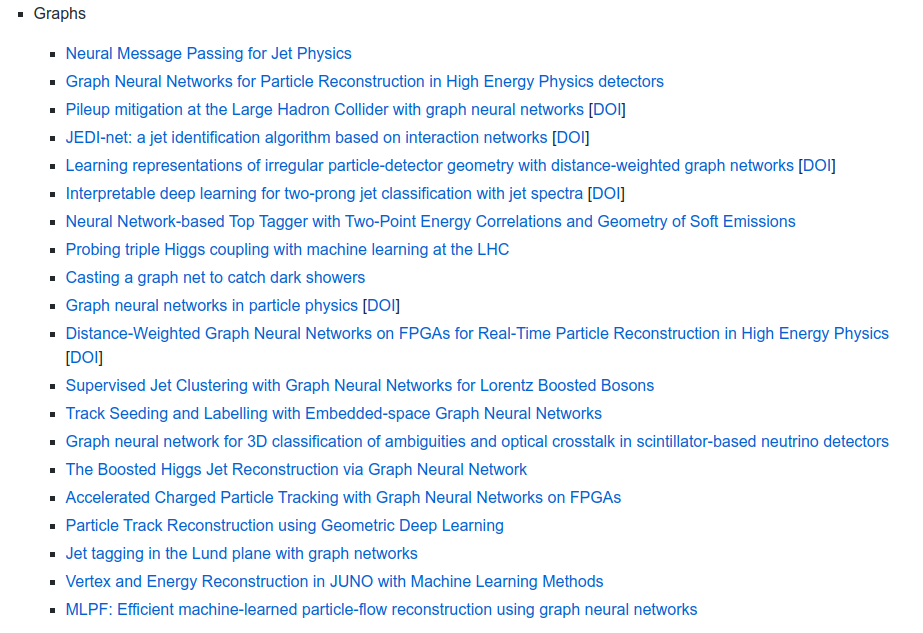}}
    \caption{A snapshot of the Markdown website form of the review, with topic papers hyperlinked to their references and, when available, DOIs.}
    \label{fig:web_view}
\end{figure}

The \textit{Living Review} (\url{https://github.com/iml-wg/HEPML-LivingReview}) also includes a list of ``normal'' (unchanging) reviews within.  The remainder of the references are organized into a small number of topics to make searching through them efficient for the user. Papers may be referenced in more than one category. The fact that a paper is listed in the review does not endorse or validate its content --- that is for the community (and for peer-review) to decide. Furthermore, the classification is a best attempt and may have flaws and community input is requested if (a) we have missed a paper you think should be included, (b) a paper has been misclassified, or (c) a citation for a paper is not correct or if the journal information is now available.
The review is built automatically from the contents of the Git repository on GitHub using \LaTeX{} focused continuous integration services and after passing validation checks PDF and Markdown versions, seen respectively in \cref{fig:descriptions} and \cref{fig:web_view}, are automatically deployed through continuous delivery to a web accessible area on GitHub.
In addition to providing a living PDF hosted on GitHub, we also provide a corresponding BibTeX file that anyone can use when they write new papers.
Please check back before you post your own paper to arXiv to ensure that you have the latest updates.

Note that this Living Review does not provide a review of machine learning in general.  Some methods (e.g. Generative Adversarial Networks~\cite{Goodfellow:2014upx}) have citations within the Living Review, but we encourage you to look elsewhere for original research and reviews in areas of pure and applied machine learning outside of HEP.

The purpose of this paper is to briefly introduce the structure of the Living Review (Sec.~\ref{sec:categories}) and describe how to contribute (Sec.~\ref{sec:contribute}).
The note ends with outlook (Sec.~\ref{sec:outlook}) and conclusions (Sec.~\ref{sec:concl}).
Furthermore, it will serve as an unchanging reference to the review, which may be useful in some cases.

\section{Categories}
\label{sec:categories}

\begin{figure}[h!]
    \centering
    \fbox{\includegraphics[width=\linewidth]{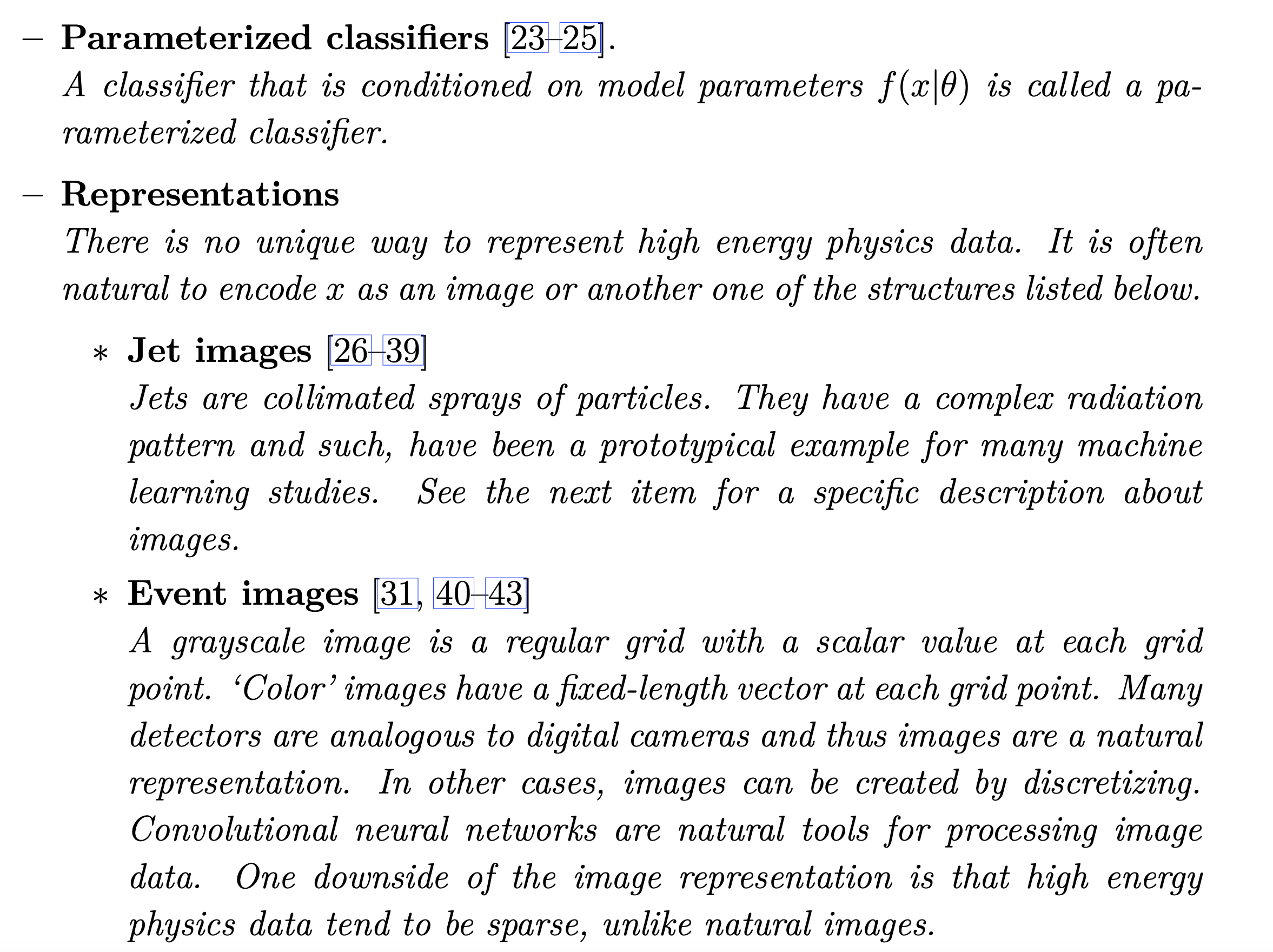}}
    \caption{A snapshot of the PDF form of the review, which includes descriptions for each category and sub-category.}
    \label{fig:descriptions}
\end{figure}

Organizing papers into topics is critical for discoverability.  Most papers do not provide keywords and often do not specify enough information in their title and/or abstract to be automatically categorized.  Therefore, we have proposed a list of categories and manually place papers into groups.  A single paper can be in more than one group.  As with all parts of the review, the categorized are alive and may change and expand as the field evolves.  Categories include Classification, Regression, Generation, Anomaly Detection, and more.  Furthermore, sub-categories are provided in some cases when there are multiple research directions within a particular category.  We have also provided brief descriptions for each category and sub-category, as illustrated in~\cref{fig:descriptions}.

\section{Contributing}
\label{sec:contribute}

\begin{figure}[h!]
    \centering
    \fbox{\includegraphics[width=\linewidth]{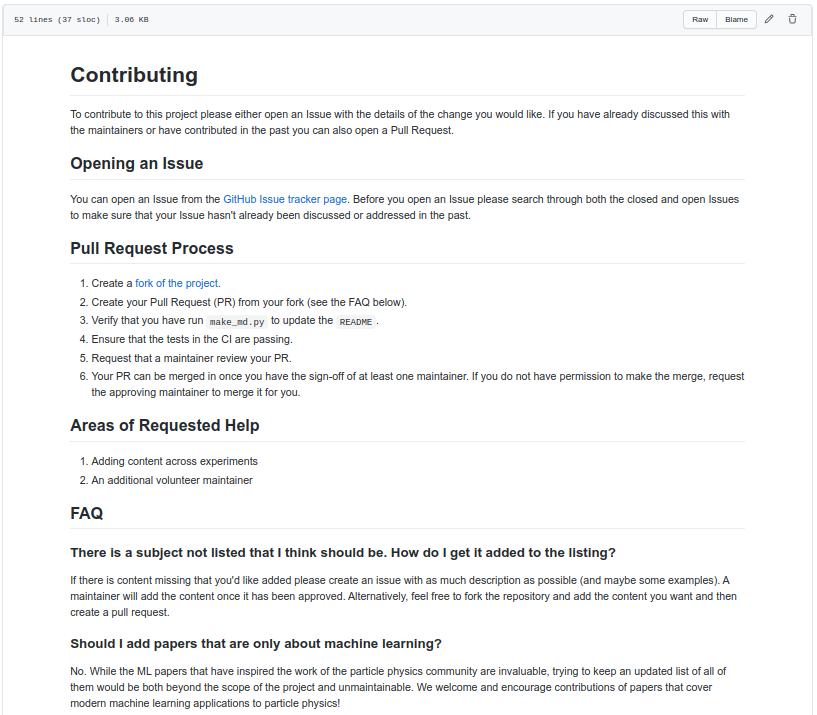}}
    \caption{A snapshot of the \texttt{CONTRIBUTING.md} document detailing the guidelines for contributions from non-maintainers to the review.}
    \label{fig:CONTRIBUTING}
\end{figure}

In addition to being a living document that is updated on demand with the release of new publications, the review also benefits from community involvement.
Anyone may --- and frequent contributors have --- submit a new paper or document to the review in the form of a contribution through a pull request (PR) to the review's \href{https://github.com/iml-wg/HEPML-LivingReview}{GitHub project}.
To help steer new contributions and ensure a smooth PR process and review with the maintainers, a contributions guide is located in the project's Git repository in the form of a ``\texttt{CONTRIBUTING.md}'' document, seen partially in \cref{fig:CONTRIBUTING} --- a project staple in the Open Source community.
The contributing guide gives detailed instruction and examples on the recommended procedures and software workflow to make revisions and additions, and additionally addresses frequently asked questions new contributors may have.

\section{Future Plans}
\label{sec:outlook}

Suggestions for new features can be submitted to the Living Review by creating a \href{https://github.com/iml-wg/HEPML-LivingReview/issues}{GitHub issue} as documented in the project's \texttt{CONTRIBUTING.md}.  There are already several key features that we would like to add in the future, mostly related with various levels of automation.  The most basic update we want to add is to automatically update paper references.  Papers are mostly added to the Living Review when they are posted to arXiv.   Journal references are currently only added in an ad-hoc fashion.   One way this could be implemented is to be synched with \href{https://inspirehep.net}{Inspire} following the links from the preprints.  This will not work for all papers, as they are not all listed on arXiv and may not be listed on Inspire.  The longer term vision is for some parts of the daily update to be automated.  It seems unlikely that this can be completely automated given the rapidly changing nature of the field (and thus what constitutes the field), but certainly quires for some key words may be able to catch a significant fraction of new papers posted to arXiv.

\section{Conclusions}
\label{sec:concl}

This paper has described the Living Review of Machine Learning for High Energy Physics.  The review will continuously evolve as new papers are written in this area and we welcome and encourage community contributions to any aspect of the project.  Machine learning holds great potential to significantly enhance the way we do HEP, both experimentally and theoretically, as is becoming well-documented by the growing literature in this area.  We hope that the Living Review is a useful tool to keep track of this rapid progress.

\begin{acknowledgments}

We are grateful to the CERN Inter-Experimental LHC Machine Learning Working Group (IML) for supporting the Living Review initiative.
We would like to particularly thank Loukas Gouskos, David Rousseau, Pietro Vischia, and Riccardo Torre who helped us define the scope of the Living Review project and have graciously agreed to allow the review to be hosted on the IML GitHub.
We are also grateful to everyone in the HEP community has contributed to the review.
We would also like to thank Martin Erdmann and Kyle Cranmer for their support and encouragement at the beginning of this project.
BN is supported by the Department of Energy, Office of Science under contract number DE-AC02-05CH11231.
MF is supported in part by the National Science Foundation, under cooperative agreement OAC-1836650.

\end{acknowledgments}


\bibliography{bib/ref}

\end{document}